\documentclass[]{pasj02} 
\usepackage[switch,mathlines]{lineno} 

\jyear{2026}
\Received{}
\Accepted{}

\begin{document} 

\title{Kinetic structure of the intracluster medium across nearby clusters observed with XRISM }

\author{
Naomi \textsc{Ota},\altaffilmark{1,2}\altemailmark\orcid{0000-0002-2784-3652} \email{naomi@cc.nara-wu.ac.jp} 
Erwin T. \textsc{Lau},\altaffilmark{1}\orcid{0000-0001-8914-8885}
Satoshi \textsc{Yamada},\altaffilmark{3,4,5}\orcid{0000-0002-9754-3081}
Yuki \textsc{Omiya},\altaffilmark{6}\orcid{0009-0009-9196-4174}
and
Hiroya \textsc{Yamaguchi}\altaffilmark{6}\orcid{0000-0002-5092-6085}
}

\altaffiltext{1}{Department of Physics, Nara Women's University, Kitauoyanishi-machi, Nara, Nara 630-8506, Japan}
\altaffiltext{2}{Argelander-Institut f\"{u}r Astronomie (AIfA), Universit\"{a}t Bonn, Auf dem H\"{u}gel 71, 53121 Bonn, Germany}
\altaffiltext{3}{Frontier Research Institute for Interdisciplinary Sciences, Tohoku University, Sendai, Miyagi 980-8578, Japan}
\altaffiltext{4}{Department of Astronomy, University of Geneva, Ch. d’Ecogia 16, 1290, Versoix, Switzerland.}
\altaffiltext{5}{Astronomical Institute, Tohoku University, 6-3 Aramakiazaaoba, Aoba-ku, Sendai, Miyagi 980-8578, Japan}
\altaffiltext{6}{Institute of Space and Astronautical Science (ISAS), Japan Aerospace Exploration Agency (JAXA), Kanagawa 252-5210, Japan}



\maketitle

\begin{abstract}
XRISM/Resolve is building a sample of galaxy clusters with directly measured ICM gas motions, revealing diverse projected dynamical states. We compile 45 XRISM/Resolve measurements in 19 nearby galaxy clusters and place them on a common, emission-weighted effective line-of-sight scale, $\ell_{\rm eff}$. We compare the line-of-sight velocity dispersion $\sigma_v$, bulk velocity amplitude $|v_{\rm bulk}|$, their ratio $R_v \equiv |v_{\rm bulk}|/\sigma_v$, and non-thermal pressure proxies. Disturbed non-cool-core systems are not simply higher-dispersion counterparts of relaxed cool-core regions. Instead, differences among cool-core centers, cool-core outer regions, and non-cool-core systems are driven mainly by coherent line-of-sight motion relative to unresolved line broadening: $R_v$ tends to remain below unity in cool-core regions but often exceeds unity in non-cool-core systems, with the mean $R_v$ rising from $0.45$ in cool-core centers to $1.6$ in non-cool-core systems. These diagnostics help separate local central line broadening, likely associated with AGN feedback in some cool cores, from larger-scale coherent motions associated with sloshing, mergers, and halo assembly. Comparison with forward-modeled TNG-Cluster predictions suggests that many cool-core measurements occupy the lower part of the predicted non-thermal pressure range, consistent with small hydrostatic-mass corrections in relaxed systems and larger corrections in disturbed ones. XRISM is thus beginning to resolve the projected kinetic structure of the ICM across cluster environments, rather than tracing a single sequence of increasing turbulence.
\end{abstract}


\section{Introduction}
\label{sec:intro}

Galaxy clusters grow hierarchically through mergers and continuous accretion, converting gravitational energy into thermal and kinetic energy of the intracluster medium (ICM). As gas falls into the dark-matter potential, it is shock heated to X-ray--emitting temperatures, while the same assembly process drives bulk flows and turbulence. In cluster cores, gas sloshing and energy injection from central active galactic nuclei (AGN) further perturb the gas through jets, bubbles, shocks, sound waves, and uplift. The ICM velocity field therefore provides a direct dynamical view of how hot baryons respond to halo assembly and core evolution \citep{Lau09}.

Gas motions also affect cluster mass measurements. Their non-thermal pressure supplements thermal pressure support and can bias hydrostatic mass estimates, a major systematic uncertainty in cluster cosmology \citep{Ota18,Pratt19}. Identifying where such support is small enough for hydrostatic equilibrium to be reliable, and where coherent motions make greater caution necessary, is therefore a key goal for both ICM physics and mass calibration.

High-resolution X-ray spectroscopy with XRISM/Resolve \citep{Tashiro25,Ishisaki25,Kelley25} has begun to deliver a growing set of direct measurements of ICM gas motions through Doppler shifts and line broadening of the Fe--K complex. Published XRISM/Resolve cluster measurements now cover a range of dynamical states, motivating a simple diagnostic framework for comparing coherent and unresolved ICM gas motions. In this Letter, we use the velocity ratio
\begin{equation}
R_v \equiv |v_{\rm bulk}|/\sigma_v
\end{equation}
as a phenomenological measure of whether the projected velocity structure is dominated by centroid shifts or by line broadening.

We compile XRISM/Resolve measurements for 19 galaxy clusters, comprising 45 pointings and analysis regions, and compare $\sigma_v$, $|v_{\rm bulk}|$, $R_v$, and non-thermal pressure proxies in a common framework. The velocity ratio $R_v$ has been used to interpret gas motions in individual XRISM systems \citep{Zhang26b,Ota26}; here we apply it to the full current compilation. 

A central aim is to place these diagnostics on a common aperture scale. Because the measurements sample different physical volumes and projection depths depending on cluster distance and surface-brightness profile, we compare them using an emission-weighted effective line-of-sight length, $\ell_{\rm eff}$ (Section~\ref{sec:sample}), and examine how the observed gas motions are shaped by both aperture scale and dynamical state. We use the cool-core/non-cool-core classification as a practical organizing label rather than a strict evolutionary sequence, and compare the emerging trends with TNG-Cluster cosmological simulation  \citep{Nelson24}.

Throughout this Letter, we adopt a flat $\Lambda$CDM cosmology with $H_0 = 70~\mathrm{km\,s^{-1}\,Mpc^{-1}}$, $\Omega_{\rm m} = 0.3$, and $\Omega_\Lambda = 0.7$. Unless otherwise stated, quoted uncertainties correspond to the 68\% confidence level.

\section{XRISM sample and velocity diagnostics}
\label{sec:sample}

Table~\ref{tab:xrism_sample} lists the XRISM/Resolve measurements used in this work, covering 45 pointings and analysis regions in 19 nearby galaxy clusters. We include measurements from published papers or preprints when region-level Fe--K velocities and the information needed to estimate the effective line-of-sight scale are reported. For each pointing or region, we adopt the line-of-sight velocity dispersion $\sigma_v$, measured from the intrinsic width of the Fe--K complex, and the line-of-sight bulk velocity $v_{\rm bulk}=c(z_{\rm ICM}-z_{\rm ref})/(1+z_{\rm ref})$, where $z_{\rm ICM}$ is the Fe--K line-centroid redshift and $z_{\rm ref}$ is the reference redshift adopted in the source paper. In most systems, $z_{\rm ref}$ is the redshift of the brightest cluster galaxy (BCG) or central galaxy; in strongly disturbed mergers such as Coma, A1914, and A2034, where a unique BCG frame is not well defined, we retain the mean member-galaxy redshift used in the source papers.

To compare measurements obtained with different angular apertures and redshifts, we assign each region an effective length scale $\ell_{\rm eff}$, following the effective line-of-sight depth concept of \citet{Zhuravleva12}. We model the gas density profile with a single- or double-$\beta$ form, using mainly published ACCEPT \citep{Donahue26,Cavagnolo09} and HIFLUGCS \citep{Eckert11} profiles, supplemented by dedicated literature profiles when available. Assuming emissivity proportional to density squared, we integrate the emission measure along the line of sight out to $R_{200}$, where the mean enclosed density is 200 times the critical density. At each projected radius, the local effective length is the half-length of the sky-plane-centered line-of-sight interval containing 50\% of the emission measure, averaged over the extraction region with projected-emission-measure weighting. While this averaging accounts for spatial variations in the line-of-sight depth across the extraction region, the transverse size of the aperture itself is not included in the definition of $\ell_{\rm eff}$. Thus, $\ell_{\rm eff}$ is an emission-weighted characteristic line-of-sight scale sampled by Resolve, not a direct turbulence injection scale.

The same observational caution applies to the velocity diagnostics. Both $\sigma_v$ and $|v_{\rm bulk}|$ are emissivity-weighted, projected aperture quantities affected by projection, point-spread-function mixing, and aperture averaging. We therefore interpret $\sigma_v$ not as turbulence alone, but as unresolved velocity structure within the aperture, while $|v_{\rm bulk}|$ represents the aperture-averaged centroid shift relative to the adopted reference frame.

We quantify the kinetic importance of gas motions using two non-thermal pressure proxies, following the spatially resolved XRISM analysis of A2029 \citep{XRISM25b}. The sound speed is $c_s=(\gamma k_{\rm B}T/\mu m_{\rm p})^{1/2}$, computed from the best-fit spectroscopic temperature with $\gamma=5/3$ and $\mu=0.61$. If line broadening is interpreted as isotropic random motion, the three-dimensional Mach number is $\mathcal{M}_{\rm 3D}=\sqrt{3}\sigma_v/c_s$, and the corresponding turbulent pressure proxy is $\alpha_{\rm turb}\equiv P_{\rm nt,turb}/P_{\rm tot}=\mathcal{M}_{\rm 3D}^2/(\mathcal{M}_{\rm 3D}^2+3/\gamma)$. Including coherent line-of-sight bulk motion, we define $\mathcal{M}_{\rm 3D,eff}=(3\sigma_v^2+v_{\rm bulk}^2)^{1/2}/c_s$ and $\alpha\equiv P_{\rm nt}/P_{\rm tot}=\mathcal{M}_{\rm 3D,eff}^2/(\mathcal{M}_{\rm 3D,eff}^2+3/\gamma)$. In the generalized hydrostatic equilibrium equation, non-thermal support enters through the gas kinetic stress tensor, which includes both random and coherent velocity components \citep{Lau13}; together with its radial gradient, this term determines the corresponding hydrostatic mass correction. Here $\alpha_{\rm turb}$ uses $\sigma_v$ alone, whereas $\alpha$ adds $v_{\rm bulk}$ and matches the effective-3D-velocity non-thermal pressure fraction adopted in the A2029 mass-bias analysis. Evaluated per aperture, $\alpha$ should therefore be regarded as a projected estimate of this fraction, rather than a full radial deprojection of the local pressure support.

For interpretation, we assign each measurement a broad phenomenological class. Cool-core regions inside and outside the cooling radius are labeled ``CC center'' and ``CC outer'', respectively, and denoted CC/C and CC/O in Table~\ref{tab:xrism_sample}. Non-cool-core measurements are labeled NCC/C or NCC/O there, but are combined into a single NCC class in the diagnostic plots and summary statistics. This CC/NCC classification is not a strict merger classification: cool cores may be sloshing or AGN-perturbed, while most NCC systems here are disturbed or merging. Because the present sample contains only one weak-cool-core (WCC) system, A3571, we merge WCC with CC and treat the labels as phenomenological guides.

For comparison with cosmological simulations, we use the forward-modeled TNG-Cluster results of \citet{Lau26}, including all 352 simulated clusters without imposing any additional selection criteria. Hot gas cells with $T>10^{6}\,{\rm K}$ are projected along three orthogonal axes with XRISM/Resolve response weighting to obtain emission-weighted maps of $T$, $v_{\rm bulk}$, $\sigma_v$, and $\ell_{\rm eff}$. The non-thermal pressure proxies are computed with the same definitions as for the observations. We pool valid pixels from all projections, bin them by $\ell_{\rm eff}$, and show the median and 16th--84th percentile ranges for simulated CC, WCC, and NCC clusters, classified by the cooling time at $0.015R_{500c}$ as $<1.0$~Gyr, $1.0$--$7.7$~Gyr, and $\geq7.7$~Gyr, respectively. We limit $\ell_{\rm eff}\geq 10$~kpc in the simulation to minimize potential impact of numerical resolution. Further details are given in \citet{Lau26}.

\section{ICM kinetic structure in the current XRISM cluster sample}
\label{sec:results}

\begin{figure*}
\begin{center}
\includegraphics[width=0.99\textwidth]{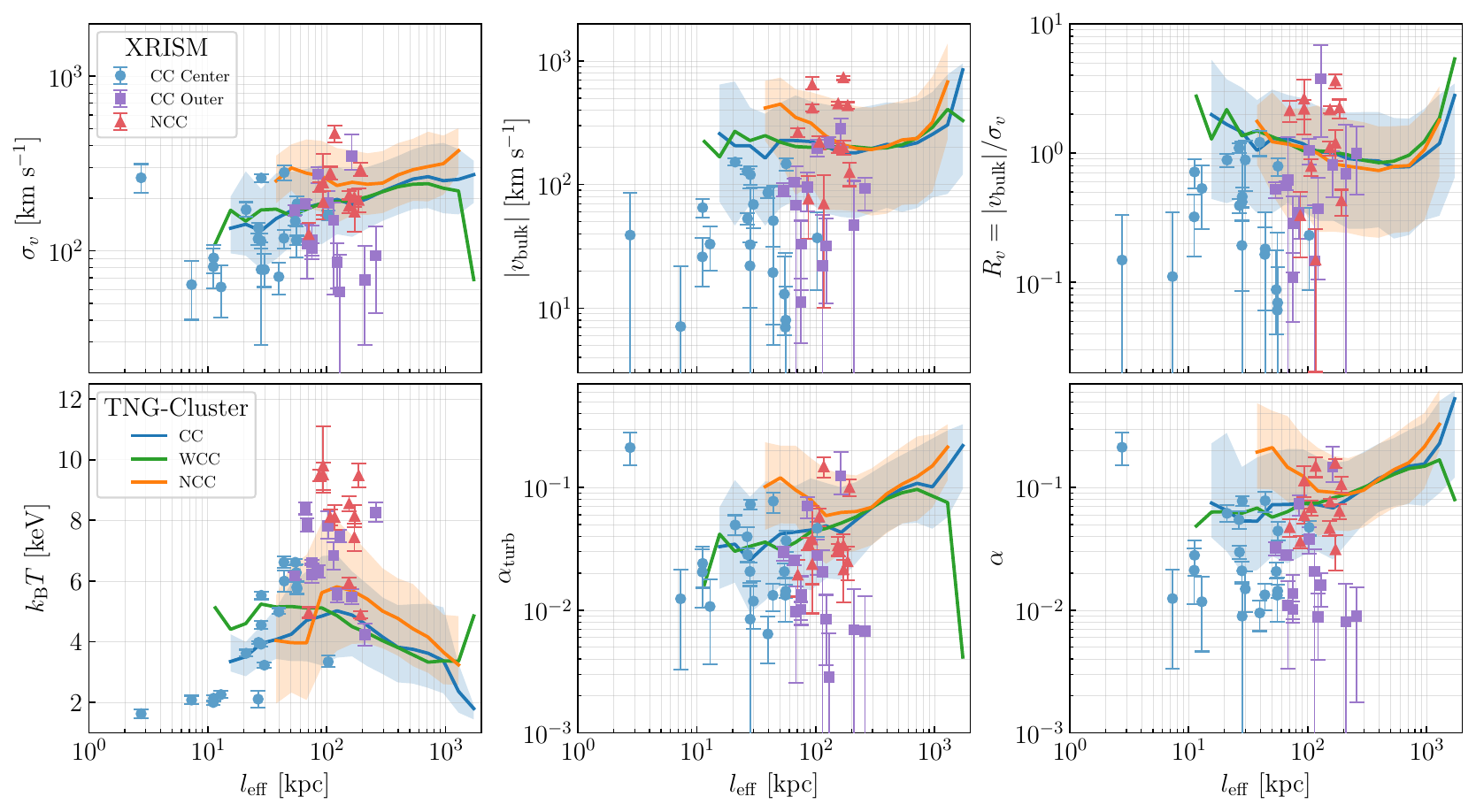}
\end{center}
\caption{XRISM velocity diagnostics versus $\ell_{\rm eff}$. The six panels show $\sigma_v$, $|v_{\rm bulk}|$, $R_v$, gas temperature, and the non-thermal pressure fractions $\alpha_{\rm turb}$ and $\alpha$, where $\alpha$ includes both line broadening and coherent line-of-sight bulk motion. Symbols show XRISM measurements, with circles, squares, and triangles denoting CC center, CC outer, and NCC regions, respectively. Curves show spectroscopically weighted TNG-Cluster profiles for simulated CC, WCC, and NCC systems in blue, green, and orange, respectively. Shaded bands indicate the 16th--84th percentile ranges for CC and NCC; the WCC median track is shown without a shaded range for reference. We set a lower limit $\ell_{\rm eff}\geq 10$~kpc in the TNG-Cluster result to minimize potential impact of numerical resolution. Because the present sample contains only one WCC system, we do not define a separate observational WCC class. Error bars show quoted statistical uncertainties. {Alt text: Six diagnostic panels plot XRISM cluster velocity measurements against effective line-of-sight length. The panels show velocity dispersion, absolute bulk velocity, velocity ratio, gas temperature, and two non-thermal pressure proxies, compared with TNG-Cluster profiles.}}\label{fig:velocity_diagnostics}
\end{figure*}

Figure~\ref{fig:velocity_diagnostics} shows the line-of-sight velocity diagnostics compiled in Table~\ref{tab:xrism_sample} as a function of the effective length scale, $\ell_{\rm eff}$, with the TNG-Cluster distributions overplotted.  The sample spans almost two orders of magnitude in $\ell_{\rm eff}$, from central cool-core regions to outer cool-core pointings and dynamically disturbed systems sampling up to a few hundred kpc along the line of sight.  The XRISM measurements are grouped into CC center, CC outer, and NCC regions to distinguish cooling state and pointing location, rather than to impose a radial or evolutionary sequence. The temperature panel is shown to provide thermodynamic context for the kinematic measurements, since $kT$ sets the sound speed entering the non-thermal pressure proxies.

The velocity dispersion $\sigma_v$ shows substantial scatter at fixed $\ell_{\rm eff}$.  Most cool-core measurements have modest line broadening, with $\sigma_v\sim100$--$200~{\rm km\,s^{-1}}$, while higher values occur in AGN-perturbed or quasar-host systems such as Virgo/M87, Cygnus A, and H1821+643, and in disturbed mergers.  The weak positive correlation with $\ell_{\rm eff}$ likely reflects sampled physical scale, local feedback, sloshing, shear, and projected multi-component structure, rather than a simple increase of local turbulence with scale.

The bulk velocity provides a complementary diagnostic.  Several relaxed-looking cool-core systems have small $|v_{\rm bulk}|$, as in A2029, where XRISM finds very small non-thermal support out to $R_{2500}$ in the observed direction \citep{XRISM25b}.  By contrast, A3395S has a low dispersion but a significant line-of-sight bulk velocity, showing that coherent motion can be important even when line broadening is modest \citep{Ota26}.  In Figure~\ref{fig:velocity_diagnostics}, high $R_v$ values occur mainly in dynamically disturbed systems.  This trend is not solely set by the reference redshift for the strongest mergers: re-referencing Coma, A1914, and A2034 to plausible BCG frames changes the NCC mean from $R_v=1.6\pm1.1$ to $R_v\simeq1.0$--$2.1$, while removing them gives $R_v\simeq1.2$.  These values remain above the CC-center mean of $0.45\pm0.36$, albeit with larger uncertainty.  Thus, large $R_v$ values provide suggestive evidence for important coherent bulk motions in many NCC and merging systems.

The two non-thermal pressure proxies summarize the same distinction: $\alpha_{\rm turb}$ is typically a few per cent in relaxed cool-core regions, whereas including coherent line-of-sight velocity raises $\alpha$ in systems with large bulk motions.  Compared with the forward-modeled TNG-Cluster distributions described in Section~\ref{sec:sample}, the XRISM points occupy the same broad region as the simulations, but tend to populate the lower part of the predicted non-thermal pressure range, not only in CC centers but also in several CC outer pointings.  By contrast, NCC pointings extend toward the simulated disturbed branch, where larger $R_v$ and higher non-thermal pressure proxies are expected.  The large scatter among CC outer regions is therefore informative: these apertures can sample quiet azimuths, sloshing sectors, or outer-core gas affected by past assembly, making this class particularly sensitive to viewing direction and aperture placement.  In such regions, low $\sigma_v$ does not exclude coherent sloshing or large-scale flow, because motions coherent across the Resolve aperture appear mainly as centroid shifts rather than line broadening.

\section{Discussion: interpreting the XRISM velocity diagnostics}
\label{sec:discussion}

We interpret the XRISM velocity diagnostics in terms of dynamical state, spatial sampling, and physical drivers. We first examine the empirical partition between line broadening and coherent centroid shifts, and then connect it to cluster relaxation, halo assembly, and central feedback.

\subsection{Velocity partition across dynamical states}
\label{subsec:dynstate}

The clearest quantitative result is that disturbed systems are not simply higher-dispersion versions of relaxed cool cores. When binned by dynamical state (Table~\ref{tab:summary_stats}), the mean $|v_{\rm bulk}|$ increases from CC Center to NCC much more strongly than $\langle\sigma_v\rangle$, while the turbulent Mach number $\mathcal{M}_{\rm 3D}$ remains subsonic and nearly constant across bins because NCC systems are hotter. Over the same comparison, the total non-thermal pressure fraction $\alpha$ increases, whereas $\alpha_{\rm turb}$ remains comparable within the scatter and neither pressure proxy is monotonic through the intermediate CC Outer bin. The state dependence is therefore driven mainly by the growing coherent velocity component relative to line broadening, not by a uniform increase in the Mach-scaled strength of unresolved motions.

\begin{figure}
\begin{center}
\includegraphics[width=0.45\textwidth]{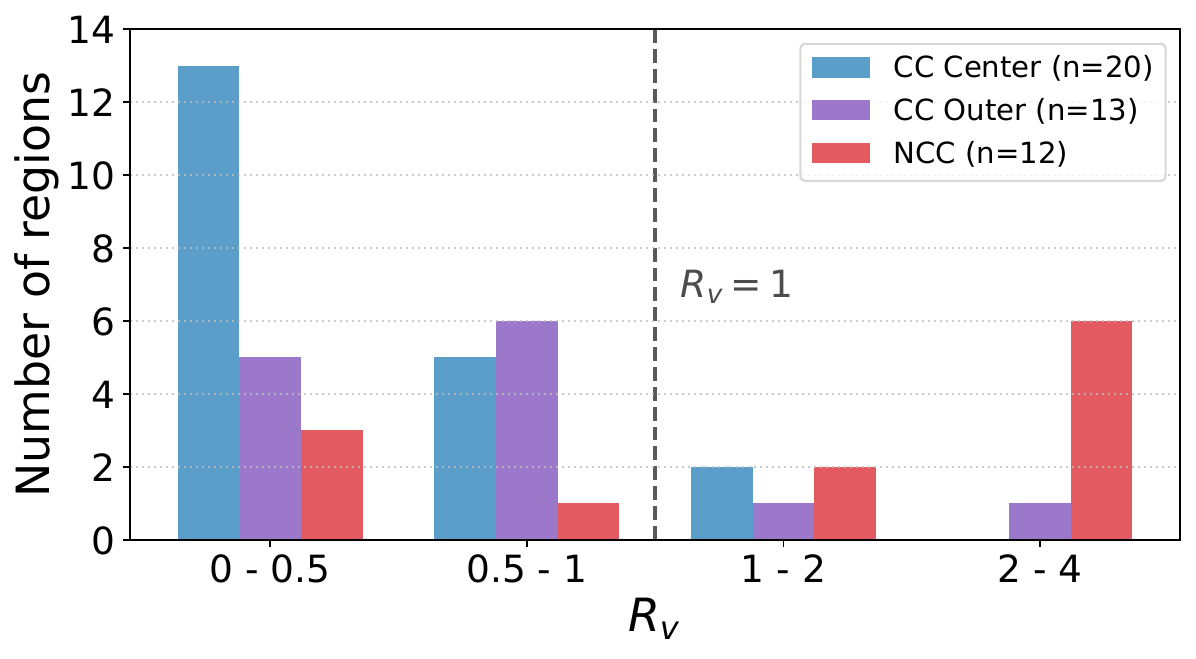}
\end{center}
\caption{
Distribution of the velocity ratio $R_v = |v_{\rm bulk}|/\sigma_v$ for the XRISM region-level sample, separated into CC Center, CC Outer, and NCC regions. The dashed vertical line marks $R_v=1$, where the coherent bulk velocity equals the line-of-sight velocity dispersion. Most cool-core regions are concentrated below unity, with one CC outer region extending to high $R_v$, whereas NCC regions extend more often above unity, indicating that coherent motions are relatively more important in dynamically disturbed systems. {Alt text: Three histograms compare the velocity ratio for CC Center, CC Outer, and NCC regions. CC Center values mostly lie below one, while NCC values extend more often above one.}}
\label{fig:vbulk_sigma_hist}
\end{figure}

This interpretation is supported by $R_v$ and by suggestive scale trends. Figure~\ref{fig:vbulk_sigma_hist} shows that the $R_v$ distribution shifts systematically from CC Center to CC Outer and then to NCC; most CC Center regions lie below unity, whereas many NCC regions exceed unity. In the region-level sample, both $\sigma_v$ and $|v_{\rm bulk}|$ increase with $\ell_{\rm eff}$, with a stronger rank correlation for $|v_{\rm bulk}|$ than for $\sigma_v$ ($\rho=0.48$, $p=8.0\times10^{-4}$, and $\rho=0.32$, $p=3.3\times10^{-2}$, respectively). Here $\rho$ is Spearman's rank coefficient and $p$ is the corresponding two-sided probability. Because several clusters contribute multiple regions, these formal significances are likely optimistic; cluster-level resampling and cluster--class averaging weaken the correlations, especially for $\sigma_v$. We therefore treat the $\ell_{\rm eff}$ trends as suggestive evidence that larger apertures preferentially capture sloshing, projection, and large-scale coherent motions, rather than as precise scaling relations.

We also examined whether these trends could be driven by the broad halo-mass range of the sample. Using $M_{500}$ and $R_{500}$ estimates primarily from the MCXC-II catalog \citep{Sadibekova24}, we define $V_{500}=(GM_{500}/R_{500})^{1/2}$ and normalize the measured velocities by this characteristic halo velocity. The correlation of $\sigma_v/V_{500}$ with $\ell_{\rm eff}$ becomes weaker and is no longer statistically significant ($\rho=0.17$, $p=0.26$), whereas $|v_{\rm bulk}|/V_{500}$ retains a significant positive correlation with $\ell_{\rm eff}$ ($\rho=0.39$, $p=7.5\times10^{-3}$). The dynamical-state dependence shows the same qualitative behavior: $|v_{\rm bulk}|/V_{500}$ increases systematically from CC Center to CC Outer and NCC, while $\sigma_v/V_{500}$ shows a weaker progression. Since $R_v=|v_{\rm bulk}|/\sigma_v$ is unchanged by the $V_{500}$ normalization, these results indicate that halo-mass scaling may contribute to the weak $\sigma_v$--$\ell_{\rm eff}$ trend, but cannot account for the increasing relative importance of coherent motion toward larger-scale and dynamically disturbed regions.

As an additional check, we examined the direct dependence on $M_{500}$ using one median value per cluster to avoid overweighting systems with multiple regions. The $M_{500}$--$\sigma_v$ relation shows a weak positive trend broadly consistent with the expected $\sigma_v\propto M_{500}^{1/3}$ scaling, but the correlation is not statistically significant ($\rho=0.19$, $p=0.45$). We likewise find no significant correlations of $M_{500}$ with $|v_{\rm bulk}|$ or $R_v$, with $\rho=-0.03$ and $-0.07$ ($p=0.90$ and $0.79$), respectively. Given the heterogeneous radial and spatial coverage of the XRISM measurements, we regard this comparison as a check for a dominant mass-driven trend rather than a precise determination of the mass--velocity scaling relation.

The effective length scale introduces an important caveat, because aperture scale and dynamical state are partly coupled. The mean $\ell_{\rm eff}$ increases from CC Center to CC Outer and NCC (Table~\ref{tab:summary_stats}). However, scale alone does not describe the velocity partition: while CC Outer and NCC regions have comparable mean $\ell_{\rm eff}$, NCC regions show substantially larger mean $|v_{\rm bulk}|$ and $R_v$. We therefore interpret the CC Center--CC Outer--NCC progression as an empirical scale--dynamical state trend in projected velocity structure: differences in the sampled scale contribute to the scatter in the velocities, while the increasing dominance of coherent motion reflects dynamical state.

As a check against overweighting clusters with multiple regions, we repeated the summary statistics after averaging measurements within each cluster and class; the same qualitative trends are recovered, with NCC systems retaining the largest $|v_{\rm bulk}|$, $R_v$, and $\alpha$.

The low $\alpha_{\rm turb}$ and $\alpha$ values in many CC Center and CC Outer regions imply small hydrostatic mass corrections on the aperture scales sampled by XRISM, typically only a few per cent if $P_{\rm nt}/P_{\rm tot}$ varies slowly with radius. Some NCC regions instead reach $\alpha\gtrsim10\%$, indicating potentially appreciable departures from hydrostatic equilibrium; the exact bias depends on radial gradients, anisotropy, and whether coherent motions act as pressure support.

\subsection{From ICM kinetic structure to cluster assembly}
\label{subsec:assembly}

The scale--dynamical state trend identified above is broadly consistent with a relaxation picture. In TNG-Cluster, \citet{Lau26} find larger non-thermal support in late-forming clusters and weaker gas motions in early-forming systems at fixed radius. The low $\sigma_v$, $|v_{\rm bulk}|$, and $R_v$ values of many CC Center regions are therefore consistent with dynamically old cores where merger-driven motions have largely decayed, whereas NCC systems trace recent or ongoing assembly. CC Outer regions may form an intermediate class, but the comparison with TNG-Cluster also highlights a possible quiet-end mismatch: several observed CC outer pointings have low $\sigma_v$ and $\alpha_{\rm turb}$, and several CC centers have very small $|v_{\rm bulk}|$. If confirmed, this would suggest that simulations retain too much unresolved outer-core motion and possibly too much coherent core motion.

AGN feedback provides a more local channel for driving ICM motions in cool-core centers. Jets and uplift can generate turbulence and velocity shear on scales comparable to or smaller than the Resolve aperture, as seen in recent XRISM results for Perseus and M87 \citep{XRISM26b,XRISM26a}. Quasar-host systems such as H1821+643 may add a radiation-driven feedback channel distinct from radio-mode jets \citep{Yamada26}. These local components mainly affect $\sigma_v$, whereas the high-$R_v$ tail is dominated by systems with sloshing, mergers, or large-scale coherent flows. The quasar-host systems remain below unity in $R_v$ \citep{Yamada26}, further suggesting that radiative feedback alone does not produce the high-$R_v$ tail in the current sample.

Taken together, the XRISM diagnostics connect the projected gas-motion field to ICM relaxation and halo assembly. This view is analogous in spirit to cluster scaling spaces such as the X-ray fundamental plane \citep{Ota06,Fujita18}, and provides a compact way to compare the growing XRISM sample with simulations.

\section{Summary}
\label{sec:summary}

We have compiled 45 XRISM/Resolve measurements of ICM gas motions in 19 nearby galaxy clusters and compared $\sigma_v$, $|v_{\rm bulk}|$, $R_v=|v_{\rm bulk}|/\sigma_v$, and non-thermal pressure proxies on a common, emission-weighted $\ell_{\rm eff}$ scale. XRISM reveals a projected kinetic structure of the ICM rather than a single turbulence sequence. Relaxed cool-core centers generally show modest line broadening, small coherent velocity shifts, and low turbulent pressure support, whereas disturbed and merging systems extend toward larger $R_v$ and moderately higher non-thermal pressure proxies. This transition is driven mainly by the growing importance of coherent line-of-sight motion relative to unresolved velocity dispersion.

The CC Center--CC Outer--NCC progression reflects both aperture scale and dynamical state, with increasing $\ell_{\rm eff}$ and stronger coherent motions relative to line broadening. Comparison with forward-modeled TNG-Cluster predictions suggests that the XRISM measurements, especially in cool-core centers and several cool-core outer regions, occupy the lower part of the predicted non-thermal pressure range. The small non-thermal pressure proxies in several relaxed cool cores imply modest hydrostatic mass corrections, whereas larger coherent motions in disturbed systems highlight where hydrostatic mass estimates and turbulence-only interpretations of the line width require greater care. Thus, $R_v$, together with non-thermal pressure proxies, provides a compact diagnostic of projected ICM kinematic state across core and dynamical environments. Larger, uniformly selected samples from XRISM and future high-resolution X-ray missions, with matched-$\ell_{\rm eff}$ coverage and homogeneous AGN-power estimates, will be needed to test whether the $R_v$--$\alpha$ behavior defines a robust physical organization of ICM kinematic states.

\begin{table*}
\centering
\scriptsize
\tbl{Region-level ICM velocity diagnostics and non-thermal pressure proxies from XRISM/Resolve.}{
\begin{tabular}{llllllllllll}
\hline
Cluster &
Region &
$z_{\rm ref}$ &
Class &
$\ell_{\rm eff}$ &
$\sigma_v$ &
$v_{\rm bulk}$ &
$R_v$ &
$kT$ &
$\alpha_{\rm turb}$ &
$\alpha$ &
Ref. \\
 & & & &
(kpc) &
(km s$^{-1}$) &
(km s$^{-1}$) &
 &
(keV) &
(\%) &
(\%) & \\
\hline
Virgo & Reg1 & 0.00428 & CC/C & $3_{-1}^{+1}$ & $262_{-46}^{+52}$ & $+39_{-47}^{+47}$ & $0.15_{-0.18}^{+0.18}$ & $1.63_{-0.14}^{+0.15}$ & $21.1_{-6.1}^{+6.8}$ & $21.3_{-6.0}^{+6.8}$ & A \\
Virgo & Reg2 & 0.00428 & CC/C & $7_{-2}^{+2}$ & $64_{-24}^{+24}$ & $-7_{-15}^{+15}$ & $0.11_{-0.24}^{+0.24}$ & $2.08_{-0.12}^{+0.14}$ & $1.2_{-0.9}^{+0.9}$ & $1.2_{-0.9}^{+0.9}$ & A \\
Virgo & Reg3 & 0.00428 & CC/C & $13_{-3}^{+4}$ & $62_{-21}^{+21}$ & $-33_{-13}^{+13}$ & $0.53_{-0.27}^{+0.27}$ & $2.26_{-0.11}^{+0.12}$ & $1.1_{-0.7}^{+0.7}$ & $1.2_{-0.7}^{+0.7}$ & A \\
Virgo & E & 0.00428 & CC/C & $11_{-3}^{+3}$ & $81_{-20}^{+22}$ & $-26_{-11}^{+11}$ & $0.32_{-0.16}^{+0.16}$ & $2.00_{-0.07}^{+0.07}$ & $2.0_{-1.0}^{+1.1}$ & $2.1_{-1.0}^{+1.1}$ & B \\
Virgo & SW & 0.00428 & CC/C & $11_{-3}^{+3}$ & $91_{-17}^{+17}$ & $-65_{-11}^{+11}$ & $0.71_{-0.18}^{+0.18}$ & $2.14_{-0.09}^{+0.09}$ & $2.4_{-0.9}^{+0.9}$ & $2.8_{-0.9}^{+0.9}$ & B \\
Centaurus & Center & 0.01003 & CC/C & $27_{-9}^{+13}$ & $117_{-9}^{+9}$ & $-128_{-6}^{+6}$ & $1.09_{-0.10}^{+0.10}$ & $2.11_{-0.28}^{+0.28}$ & $4.0_{-0.8}^{+0.8}$ & $5.5_{-0.9}^{+0.9}$ & C,D \\
Perseus & Reg1 & 0.017284 & CC/C & $21_{-5}^{+6}$ & $172_{-16}^{+18}$ & $+152_{-10}^{+10}$ & $0.88_{-0.11}^{+0.10}$ & $3.62_{-0.11}^{+0.12}$ & $4.9_{-0.9}^{+1.0}$ & $6.2_{-0.9}^{+1.0}$ & E \\
Perseus & Reg2 & 0.017284 & CC/C & $27_{-6}^{+8}$ & $135_{-9}^{+9}$ & $+53_{-6}^{+6}$ & $0.39_{-0.05}^{+0.05}$ & $3.98_{-0.05}^{+0.05}$ & $2.8_{-0.4}^{+0.4}$ & $3.0_{-0.4}^{+0.4}$ & E \\
Perseus & Reg3 & 0.017284 & CC/C & $40_{-9}^{+11}$ & $71_{-15}^{+14}$ & $-86_{-8}^{+8}$ & $1.21_{-0.26}^{+0.28}$ & $4.99_{-0.06}^{+0.12}$ & $0.6_{-0.3}^{+0.3}$ & $0.9_{-0.3}^{+0.3}$ & E \\
Perseus & Reg4 & 0.017284 & CC/C & $56_{-13}^{+16}$ & $186_{-18}^{+20}$ & $-147_{-17}^{+18}$ & $0.79_{-0.12}^{+0.12}$ & $5.75_{-0.18}^{+0.21}$ & $3.7_{-0.7}^{+0.8}$ & $4.4_{-0.7}^{+0.8}$ & E \\
Perseus & Reg5 & 0.017284 & CC/O & $75_{-18}^{+21}$ & $115_{-25}^{+24}$ & $-33_{-18}^{+19}$ & $0.29_{-0.17}^{+0.18}$ & $6.22_{-0.27}^{+0.35}$ & $1.3_{-0.6}^{+0.6}$ & $1.4_{-0.6}^{+0.6}$ & E \\
Perseus & Reg6 & 0.017284 & CC/O & $103_{-24}^{+29}$ & $188_{-29}^{+32}$ & $+197_{-29}^{+32}$ & $1.05_{-0.24}^{+0.23}$ & $7.82_{-0.42}^{+0.48}$ & $2.8_{-0.9}^{+0.9}$ & $3.8_{-0.9}^{+1.0}$ & E \\
Perseus & E+NE & 0.017628 & CC/O & $163_{-39}^{+46}$ & $349_{-57}^{+115}$ & $+282_{-80}^{+62}$ & $0.81_{-0.35}^{+0.22}$ & $5.47_{-0.23}^{+0.26}$ & $12.4_{-3.6}^{+7.2}$ & $14.7_{-3.7}^{+6.9}$ & F \\
Perseus & N & 0.017628 & CC/O & $114_{-27}^{+32}$ & $150_{-94}^{+39}$ & $+22_{-61}^{+35}$ & $0.15_{-0.41}^{+0.25}$ & $6.84_{-0.46}^{+0.43}$ & $2.1_{-2.5}^{+1.1}$ & $2.1_{-2.5}^{+1.1}$ & F \\
Perseus & M3 & 0.017628 & CC/O & $54_{-13}^{+15}$ & $171_{-12}^{+11}$ & $-90_{-12}^{+10}$ & $0.53_{-0.08}^{+0.07}$ & $6.18_{-0.11}^{+0.12}$ & $2.9_{-0.4}^{+0.4}$ & $3.2_{-0.4}^{+0.4}$ & F \\
Perseus & O3 & 0.017628 & CC/O & $85_{-20}^{+24}$ & $275_{-30}^{+26}$ & $+95_{-34}^{+30}$ & $0.35_{-0.13}^{+0.12}$ & $6.32_{-0.20}^{+0.20}$ & $7.1_{-1.5}^{+1.3}$ & $7.3_{-1.5}^{+1.3}$ & F \\
Coma & Center & 0.02333 & NCC/C & $155_{-36}^{+43}$ & $208_{-12}^{+12}$ & $-450_{-15}^{+15}$ & $2.16_{-0.14}^{+0.14}$ & $8.55_{-0.25}^{+0.25}$ & $3.1_{-0.4}^{+0.4}$ & $7.6_{-0.5}^{+0.5}$ & C,G \\
Coma & South & 0.02333 & NCC/O & $171_{-40}^{+48}$ & $202_{-24}^{+24}$ & $-730_{-30}^{+30}$ & $3.61_{-0.45}^{+0.45}$ & $7.44_{-0.44}^{+0.44}$ & $3.4_{-0.8}^{+0.8}$ & $15.8_{-1.3}^{+1.3}$ & C,G \\
Coma & North & 0.02333 & NCC/O & $172_{-40}^{+48}$ & $167_{-39}^{+39}$ & $-199_{-26}^{+26}$ & $1.19_{-0.32}^{+0.32}$ & $8.14_{-0.36}^{+0.36}$ & $2.1_{-1.0}^{+1.0}$ & $3.1_{-1.0}^{+1.0}$ & H \\
Ophiuchus & Inner & 0.0295 & CC/C & $56_{-16}^{+19}$ & $115_{-7}^{+7}$ & $+8_{-7}^{+7}$ & $0.07_{-0.06}^{+0.06}$ & $5.80_{-0.20}^{+0.20}$ & $1.4_{-0.2}^{+0.2}$ & $1.4_{-0.2}^{+0.2}$ & C,I \\
Ophiuchus & Outer & 0.0295 & CC/O & $66_{-16}^{+19}$ & $186_{-9}^{+9}$ & $-104_{-7}^{+7}$ & $0.56_{-0.05}^{+0.05}$ & $8.40_{-0.20}^{+0.20}$ & $2.6_{-0.2}^{+0.2}$ & $2.8_{-0.3}^{+0.3}$ & C,I \\
A496 & Center & 0.0329 & CC/C & $30_{-8}^{+10}$ & $78_{-16}^{+18}$ & $-69_{-25}^{+25}$ & $0.88_{-0.38}^{+0.37}$ & $3.23_{-0.08}^{+0.09}$ & $1.2_{-0.5}^{+0.5}$ & $1.5_{-0.5}^{+0.6}$ & J \\
A3571 & R1 & 0.0386 & CC/C & $44_{-12}^{+15}$ & $118_{-15}^{+14}$ & $+19_{-12}^{+12}$ & $0.16_{-0.10}^{+0.10}$ & $6.62_{-0.19}^{+0.19}$ & $1.3_{-0.3}^{+0.3}$ & $1.3_{-0.3}^{+0.3}$ & K \\
A3571 & R2 & 0.0386 & CC/O & $75_{-19}^{+25}$ & $103_{-9}^{+8}$ & $-11_{-6}^{+6}$ & $0.11_{-0.06}^{+0.06}$ & $6.61_{-0.09}^{+0.09}$ & $1.0_{-0.2}^{+0.2}$ & $1.0_{-0.2}^{+0.2}$ & K \\
A3395S & Center & 0.05198 & NCC/C & $71_{-17}^{+21}$ & $124_{-21}^{+20}$ & $+264_{-23}^{+22}$ & $2.13_{-0.39}^{+0.40}$ & $4.96_{-0.17}^{+0.18}$ & $1.9_{-0.6}^{+0.6}$ & $4.7_{-0.8}^{+0.7}$ & L \\
A754 & Main & 0.054291 & NCC/C & $107_{-25}^{+30}$ & $279_{-23}^{+24}$ & $+220_{-29}^{+26}$ & $0.79_{-0.12}^{+0.11}$ & $8.10_{-0.26}^{+0.27}$ & $5.8_{-0.9}^{+1.0}$ & $6.9_{-0.9}^{+1.0}$ & M \\
A754 & Sub & 0.054291 & NCC/O & $186_{-44}^{+51}$ & $195_{-29}^{+32}$ & $-436_{-28}^{+26}$ & $2.24_{-0.39}^{+0.36}$ & $9.48_{-0.38}^{+0.39}$ & $2.5_{-0.7}^{+0.8}$ & $6.4_{-0.9}^{+0.9}$ & M \\
Hydra-A & Center & 0.05434 & CC/C & $103_{-25}^{+30}$ & $160_{-11}^{+10}$ & $-37_{-23}^{+23}$ & $0.23_{-0.14}^{+0.14}$ & $3.34_{-0.17}^{+0.21}$ & $4.7_{-0.7}^{+0.6}$ & $4.7_{-0.7}^{+0.6}$ & N \\
A2319 & Full & 0.05458 & NCC/C & $87_{-21}^{+25}$ & $232_{-14}^{+14}$ & $-76_{-40}^{+40}$ & $0.33_{-0.17}^{+0.17}$ & $9.48_{-0.20}^{+0.20}$ & $3.5_{-0.4}^{+0.4}$ & $3.6_{-0.4}^{+0.4}$ & C,O \\
A3667 & Center & 0.05567 & NCC/C & $154_{-37}^{+45}$ & $180_{-16}^{+17}$ & $-198_{-19}^{+17}$ & $1.10_{-0.15}^{+0.14}$ & $5.94_{-0.16}^{+0.17}$ & $3.4_{-0.6}^{+0.6}$ & $4.6_{-0.6}^{+0.6}$ & P \\
A3667 & CF\_IN & 0.05567 & NCC/O & $192_{-46}^{+56}$ & $292_{-24}^{+26}$ & $+125_{-28}^{+25}$ & $0.43_{-0.10}^{+0.09}$ & $4.89_{-0.12}^{+0.11}$ & $10.0_{-1.5}^{+1.6}$ & $10.5_{-1.5}^{+1.6}$ & P \\
Cygnus A & Core hot & 0.05608 & CC/C & $28_{-7}^{+8}$ & $261_{-13}^{+13}$ & $+120_{-20}^{+20}$ & $0.46_{-0.08}^{+0.08}$ & $5.53_{-0.13}^{+0.13}$ & $7.3_{-0.7}^{+0.7}$ & $7.7_{-0.7}^{+0.7}$ & Q \\
A1795 & R0-0.75 & 0.063001 & CC/C & $28_{-7}^{+9}$ & $114_{-11}^{+11}$ & $+22_{-12}^{+12}$ & $0.19_{-0.11}^{+0.11}$ & $3.92_{-0.07}^{+0.07}$ & $2.1_{-0.4}^{+0.4}$ & $2.1_{-0.4}^{+0.4}$ & R \\
A1795 & R0.75-1.5 & 0.063001 & CC/C & $56_{-14}^{+18}$ & $115_{-23}^{+23}$ & $+7_{-21}^{+21}$ & $0.06_{-0.18}^{+0.18}$ & $6.28_{-0.27}^{+0.27}$ & $1.3_{-0.5}^{+0.5}$ & $1.3_{-0.5}^{+0.5}$ & R \\
A1795 & R1.5-3 & 0.063001 & CC/O & $123_{-32}^{+42}$ & $86_{-25}^{+25}$ & $-32_{-21}^{+21}$ & $0.37_{-0.27}^{+0.27}$ & $5.53_{-0.22}^{+0.22}$ & $0.8_{-0.5}^{+0.5}$ & $0.9_{-0.5}^{+0.5}$ & R \\
A1795 & R3-4.5 & 0.063001 & CC/O & $209_{-54}^{+69}$ & $68_{-39}^{+39}$ & $+47_{-60}^{+60}$ & $0.69_{-0.97}^{+0.97}$ & $4.23_{-0.37}^{+0.37}$ & $0.7_{-0.8}^{+0.8}$ & $0.8_{-0.8}^{+0.8}$ & R \\
A2029 & Center & 0.0779 & CC/C & $54_{-13}^{+17}$ & $148_{-9}^{+13}$ & $+13_{-7}^{+13}$ & $0.09_{-0.05}^{+0.09}$ & $6.62_{-0.11}^{+0.11}$ & $2.1_{-0.2}^{+0.4}$ & $2.1_{-0.2}^{+0.4}$ & C,S \\
A2029 & N1 & 0.0779 & CC/O & $129_{-31}^{+37}$ & $58_{-48}^{+37}$ & $-218_{-25}^{+16}$ & $3.76_{-2.44}^{+3.12}$ & $7.49_{-0.21}^{+0.21}$ & $0.3_{-0.5}^{+0.4}$ & $1.6_{-0.5}^{+0.4}$ & C,S \\
A2029 & N2 & 0.0779 & CC/O & $259_{-62}^{+73}$ & $94_{-50}^{+44}$ & $-93_{-20}^{+31}$ & $0.99_{-0.51}^{+0.62}$ & $8.26_{-0.31}^{+0.34}$ & $0.7_{-0.7}^{+0.6}$ & $0.9_{-0.7}^{+0.6}$ & C,S \\
PKS0745-19 & Center & 0.1028 & CC/C & $28_{-7}^{+9}$ & $78_{-49}^{+34}$ & $+33_{-60}^{+60}$ & $0.42_{-0.79}^{+0.81}$ & $4.55_{-0.13}^{+0.13}$ & $0.8_{-1.1}^{+0.7}$ & $0.9_{-1.1}^{+0.8}$ & T \\
PKS0745-19 & Outer & 0.1028 & CC/O & $68_{-16}^{+20}$ & $110_{-41}^{+33}$ & $+68_{-73}^{+73}$ & $0.62_{-0.69}^{+0.71}$ & $7.85_{-0.21}^{+0.23}$ & $1.0_{-0.7}^{+0.6}$ & $1.1_{-0.8}^{+0.6}$ & T \\
A2034 & Full & 0.1132 & NCC/C & $117_{-28}^{+34}$ & $470_{-50}^{+50}$ & $+70_{-60}^{+50}$ & $0.15_{-0.13}^{+0.11}$ & $8.10_{-0.40}^{+0.40}$ & $14.8_{-2.8}^{+2.8}$ & $14.9_{-2.8}^{+2.8}$ & U \\
A1914 & Blue & 0.168 & NCC/C & $93_{-22}^{+25}$ & $190_{-30}^{+30}$ & $+416_{-27}^{+27}$ & $2.19_{-0.37}^{+0.37}$ & $9.50_{-0.50}^{+0.40}$ & $2.4_{-0.7}^{+0.7}$ & $5.9_{-0.8}^{+0.9}$ & V \\
A1914 & Red & 0.168 & NCC/C & $93_{-22}^{+25}$ & $245_{-94}^{+113}$ & $+641_{-53}^{+100}$ & $2.62_{-1.23}^{+1.08}$ & $9.80_{-0.90}^{+1.30}$ & $3.8_{-2.8}^{+3.3}$ & $11.3_{-2.9}^{+3.7}$ & V \\
H1821+643 & Core & 0.29708 & CC/C & $44_{-11}^{+15}$ & $281_{-29}^{+26}$ & $+51_{-46}^{+46}$ & $0.18_{-0.16}^{+0.16}$ & $6.00_{-0.34}^{+0.36}$ & $7.7_{-1.5}^{+1.4}$ & $7.8_{-1.5}^{+1.4}$ & W \\
\hline
\end{tabular}}\label{tab:xrism_sample}
\begin{tabnote}
Rows are sorted by the reference redshift $z_{\rm ref}$ adopted for $v_{\rm bulk}$, taken from each source paper as either the BCG or the mean member-galaxy redshift. The tabulated $\ell_{\rm eff}$ is the one-sided, 50 per cent emission-weighted line-of-sight depth, with the lower and upper errors giving the 40 and 60 per cent depths. $kT$ is the temperature used to compute sound speeds and non-thermal pressure fractions. Uncertainties on $v_{\rm bulk}$, $\alpha_{\rm turb}$, and $\alpha$ are taken from the source papers where reported for the matching region and definition, and otherwise from first-order error propagation. In the Class column, CC/NCC denote cool-core/non-cool-core systems and C/O central/outer regions. Special cases: E+NE is the joint Perseus E and NE fit; Cygnus A refers to the hotter ICM component and is upper-limit-like for $\alpha_{\rm turb}$ and $\alpha$; H1821+643 refers to the core Fe XXV-emitting gas.
References:
A = \citet{XRISM26a};
B = \citet{Simionescu26};
C = \citet{XRISM25e};
D = \citet{XRISM25d};
E = \citet{XRISM26b};
F = \citet{Zhang26a};
G = \citet{XRISM25f};
H = \citet{Gatuzz26};
I = \citet{Fujita25};
J = \citet{Veronica26};
K = \citet{McCall26};
L = \citet{Ota26};
M = \citet{Omiya26b};
N = \citet{Rose25};
O = \citet{XRISM25g};
P = \citet{Omiya26a};
Q = \citet{Majumder26};
R = \citet{Sarkar26};
S = \citet{XRISM25b};
T = \citet{Tanaka26};
U = \citet{Heinrich26};
V = \citet{Heinrich25};
W = \citet{Yamada26}.
\end{tabnote}\end{table*}

\begin{table}
\centering
\tbl{Summary statistics of the XRISM velocity diagnostics.}{
\begin{tabular}{lccc}
\hline
Quantity & CC Center & CC Outer & NCC \\
\hline
N & 20 & 13 & 12 \\
$\ell_{\rm eff}$ (kpc) & $34 \pm 23$ & $117 \pm 61$ & $133 \pm 43$ \\
$\sigma_v$ (km s$^{-1}$) & $135 \pm 67$ & $150 \pm 85$ & $232 \pm 88$ \\
$|v_{\rm bulk}|$ (km s$^{-1}$) & $56 \pm 47$ & $99 \pm 83$ & $319 \pm 216$ \\
$R_v$ & $0.45 \pm 0.36$ & $0.79 \pm 0.94$ & $1.58 \pm 1.07$ \\
$M_{\rm 3D}$ & $0.24 \pm 0.13$ & $0.20 \pm 0.12$ & $0.28 \pm 0.11$ \\
$M_{\rm 3D,eff}$ & $0.25 \pm 0.14$ & $0.22 \pm 0.13$ & $0.38 \pm 0.11$ \\
$\alpha_{\rm turb}$ (\%) & $3.7 \pm 4.6$ & $2.7 \pm 3.4$ & $4.7 \pm 3.9$ \\
$\alpha$ (\%) & $4.0 \pm 4.6$ & $3.2 \pm 3.9$ & $8.0 \pm 4.3$ \\
\hline
\end{tabular}}
\begin{tabnote}
The statistics are computed for the XRISM pointing/region-level measurements listed in Table~\ref{tab:xrism_sample}. Values are mean $\pm$ sample standard deviation. CC Center and CC Outer denote cool-core regions inside and outside the cooling radius, and NCC combines central and outer regions in non-cool-core systems. The quantities $\alpha_{\rm turb}$ and $\alpha$ are estimated from $\sigma_v$ alone and from both $\sigma_v$ and $|v_{\rm bulk}|$, respectively.
\end{tabnote}
\label{tab:summary_stats}
\end{table}

\bibliographystyle{pasj}
\bibliography{ref.bib}

\begin{ack}
We thank the XRISM project and the mission operations team for the operation of the satellite and their support of the observations. We thank the anonymous referee for careful reading of the manuscript and constructive comments. This work was supported in part by the Fund for the Promotion of Joint International Research, JSPS KAKENHI Grant Number 25K01026, 23H00121 (NO), 26K17187 (SY), 26H02075, 23K25907 (HY).
\end{ack}

\end{document}